\documentclass[10pt]{amsart}
\usepackage{amsmath,amsthm,amssymb}
\usepackage{epsfig}
\usepackage[margin=1.5in]{geometry}

\begin{document}

%%%%%%%%%%%%%%%%%%%%%%%%%%%%%%%%%%%%%%%%%%%%%%%%%%%%%%%%%%%%%%%%%%%%%%%%%%%%%%
%  Group Notation
%%%%%%%%%%%%%%%%%%%%%%%%%%%%%%%%%%%%%%%%%%%%%%%%%%%%%%%%%%%%%%%%%%%%%%%%%%%%%%

\newcommand{\Cx} {{\mathbb C}}                             %complex
\newcommand{\R} {{\mathbb R}}                              %reals
\newcommand{\Z} {{\mathbb Z}}                              %integers

\newcommand{\ep} {{\epsilon}}                              %epsilon

\newcommand{\T} {{\tau}}                                    %triangulation
\newcommand{\tri}{{\triangle}}                            %triangle
\newcommand{\ccw}{{\alpha}}                            %counter-clockwise
\newcommand{\suchthat} {\:\: | \:\:}
\newcommand{\ore} {\ \ {\it or} \ \ }
\newcommand{\oand} {\ \ {\it and} \ \ }

%%%%%%%%%%%%%%%%%%%%%%%%%%%%%%%%%%%%%%%%%%%%%%%%%%%%%%%%%%%%%%%%%%%%%%%%%%%%%%
%
%  paper formatting
%
%%%%%%%%%%%%%%%%%%%%%%%%%%%%%%%%%%%%%%%%%%%%%%%%%%%%%%%%%%%%%%%%%%%%%%%%%%%%%%

\theoremstyle{plain}
\newtheorem{thm}{Theorem}%[section]
\newtheorem{prob}[thm]{Problem}
\newtheorem{prop}[thm]{Proposition}
\newtheorem{cor}[thm]{Corollary}
\newtheorem{lem}[thm]{Lemma}
\newtheorem{conj}[thm]{Conjecture}
\newtheorem{obs}[thm]{Observation}
\theoremstyle{definition}
\newtheorem{defn}[thm]{Definition}
\newtheorem{exmp}[thm]{Example}

\theoremstyle{remark}
\newtheorem*{rem}{Remark}
\newtheorem*{nota}{Notation}
\newtheorem*{ack}{Acknowledgments}
\numberwithin{equation}{section}

\title {Compatible Triangulations and Point Partitions by Series-Triangular
Graphs}

\thanks{Authors were partially supported by NSF grants DMS-0353634 and CARGO DMS-0310354.}

\author[J.\ Danciger]{Jeff Danciger}
\address{J.\ Danciger: University of California, Santa Barbara, CA 93106}
\email{jeffdanciger@yahoo.com}

\author[S.\ Devadoss]{Satyan L.\ Devadoss}
\address{S.\ Devadoss: Williams College, Williamstown, MA 01267}
\email{satyan.devadoss@williams.edu}

\author[D.\ Sheehy]{Don Sheehy}
\address{D.\ Sheehy: Princeton University, Princeton, NJ 08544}
\email{dsheehy@princeton.edu}

\begin{abstract}
We introduce series-triangular graph embeddings and show how to partition point
sets with them.  This result is then used to improve the upper bound on the
number of Steiner points needed to obtain compatible triangulations of point
sets.  The problem is generalized to finding compatible triangulations for more
than two point sets and we show that such triangulations can be constructed with
only a linear number of Steiner points added to each point set.
\end{abstract}


\maketitle

%%%%%%%%%%%%%%%%%%%%%%%%%%%%%%%%%%%%%%%%%%%%%%%%%%%%%%%%%%%%%%%%%%%%%%%%%%%%%%
%NOTE: The \baselineskip=15pt below is the minimum needed to make the paper
%      look decent.
%      There are many superscripts {\tilde, \overline} which cause erratic
%      spaces between sentences. \baselineskip corrects these problems.  Also
%      see the \baselineskip=12pt by References.
%%%%%%%%%%%%%%%%%%%%%%%%%%%%%%%%%%%%%%%%%%%%%%%%%%%%%%%%%%%%%%%%%%%%%%%%%%%%%%

\baselineskip=15pt

%%%%%%%%%%%%%%%%%%%%%%%%%%%%%%%%%%%%%%%%%%%%%%%%%%%%%%%%%%%%%%%%%%%%%%%%%%%%%%
%
%               Introduction
%
%%%%%%%%%%%%%%%%%%%%%%%%%%%%%%%%%%%%%%%%%%%%%%%%%%%%%%%%%%%%%%%%%%%%%%%%%%%%%%
\section{Introduction}

Given two $n$ point sets in the plane, it is an open question as to whether a
compatible triangulation exists between them. This problem was first posed by Aichholzer et al.\
\cite{aichTCT} in 2002 and in a slightly different version by
Saalfeld \cite{saal} in 1987.  Aronov, Seidel, and Souvaine \cite{aron} studied
the related problem of compatibly triangulating simple polygons and showed that
$\Omega(n^2)$ Steiner points were necessary in some cases.  An immediate
corollary of Euler's Theorem states that the number of triangles in any
triangulation of a set $S$ with $n$ points is $2n-2-|CH(S)|$, where $CH(S)$ is
the convex hull of $S$. Trivially, two triangulations must have the same number
of triangles to be compatible so it is a necessary condition for a compatible
triangulation that the point sets have the same number of extreme
points. Aichholzer et al.\ conjecture that this condition is also
sufficient. They also show that compatible triangulations, for sets obeying
these necessary conditions, are always possible if one is allowed to add extra points, called \emph{Steiner} points, to the sets. The number of Steiner
points required by Aichholzer et al.\ is equal to the
number of interior points of the set minus three.

In this paper, an improved method is given for obtaining compatible
triangulations using Steiner points. This method requires the
use of a number of Steiner points approximately equal to half the
number of points in the set.
%Further, as new methods are found for compatibly triangulating point set with a larger number, $q$, of interior points, the method below allows for extension of these results to find compatible triangulations of arbitrarily large point sets using a number of Steiner points approximately equal to the number of points in the set divided by $q$.
Though finding compatible triangulations requires the
number of convex hull points to be the same in each set, this
technique makes no assumptions about the convex hulls of the point
sets. The added points increase the radii of the point sets by a
constant factor.  The method also allows for great
control over the structure of the Steiner point triangulation.

Furthermore, the question of finding $d$-way compatible
triangulations is introduced and explored. This natural generalization of the
problem asks how $d$ distinct triangulations can be triangulated so that the
resulting triangulations are pairwise compatible.  We show that for $d>2$, there
are exist $d$ sets of $n$ points which \emph{require} Steiner points in order to
yield a compatible triangulation.
%We give simple examples of $d$ different point sets all having the same number of convex hull and interior points that do not yield a $d$ way compatible triangulation without the addition of Steiner points.
The Steiner point method for the $d = 2$ case extends
for $d > 2$, producing a technique for $d$-way compatible triangulation
using a number of added points which is surprisingly independent
of $d$.

%%%%%%%%%%%%%%%%%%%%%%%%%%%%%%%%%%%%%%%%%%%%%%%%%%%%%%%%%%%%%%%%%%%%%%%%%%%%%%
%
%                Steiner Points
%
%%%%%%%%%%%%%%%%%%%%%%%%%%%%%%%%%%%%%%%%%%%%%%%%%%%%%%%%%%%%%%%%%%%%%%%%%%%%%%
\section{Steiner Points}
\subsection{}

A \emph{triangulation} of a set $S$ of points in the plane $\R^2$, denoted by $\T_S$, is a maximal set of line segments between the points such that any pair of segments intersect at one endpoint or not at all. When we refer to the triangles in a triangulation, we mean the \emph{empty} triangles, those that do not contain any other points of $S$.  This paper will only consider point sets $S$ in \emph{general position}, where  no three points of $S$ are collinear.

Let $CH(S)$ denote the convex hull of $S$.  The points of $S$ that lie on the boundary of the convex hull are the \emph{extreme points}, and the remaining points of $S$ are called \emph{interior  points}.

%The definition leads naturally to a \emph{greedy} method for finding a triangulation.  That is, we simply start with $\T_S=\emptyset$ and repeatedly add edges as long as they do not intersect.  If we do this until no more edges can be added, we have a triangulation.  We therefore get the following lemma.

%\begin{lem}
%\label{L:greedy}
%Given a set $S$ of $n$ points in $\R^2$ and a set $A$ of nonintersecting line segments between pairs of points in $S$, there exists a triangulation $\T_S$ of $S$ such that $A \subset \T_S$.  That is, $A$ can be extended to a triangulation of $S$.
%\end{lem}

\begin{defn}
\label{d:comp1}
Given two point sets $S$ and $T$ with $|S|=|T|$, along with triangulations
$\T_S$ and $\T_T$, we say $\T_S$ and $\T_T$ are \emph{compatible} if there
exists a bijection $f:S\rightarrow T$ such that $\tri(a,b,c)$ is a
clockwise-oriented triangle of $\T_S$ if and only if $\tri(f(a),f(b), f(c))$ is
a clockwise-oriented triangle of $\T_T$.
\end{defn}

The term \emph{joint triangulation} is also used in the literature to
describe the same object.  There are variants of this definition
which omit the requirement regarding orientation, requiring only that the bijection maps empty triangles to empty triangles.  We have chosen the above definition because it is more useful for applications in computer graphics and cartography.  It is important to note that none of these definitions require only the underlying graph structures of the triangulations to be isomorphic.  However, we give a definition equivalent to Definition~\ref{d:comp1} in terms of the graph structure of the triangulations.

\begin{defn}
\label{d:comp2}
Given two point sets $S$ and $T$ along with triangulations
$\T_S$ and $\T_T$, we say $\T_S$ and $\T_T$ are \emph{compatible} if the
there exists an isomorphism between their underlying graphs that also maps
$CH(S)$ to $CH(T)$ preserving the orientation.
\end{defn}

\begin{figure}[h]
\includegraphics {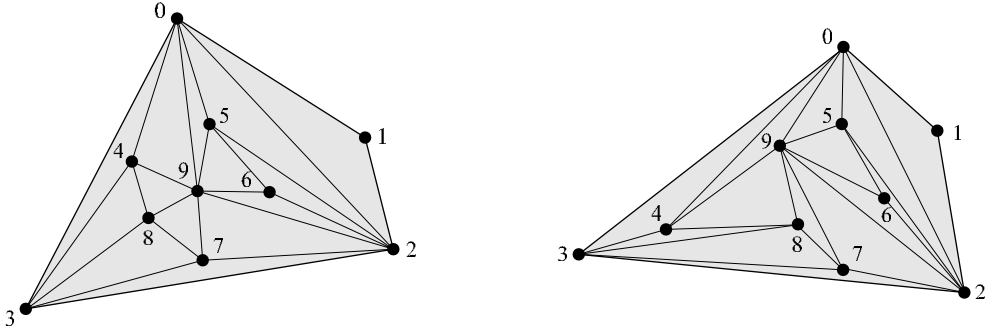}
\caption{Example of compatible triangulations.}
\label{f:compatible}
\end{figure}

\noindent Figure~\ref{f:compatible} gives an example of compatible
triangulations.  We explore the problem of finding compatible triangulations
between a given pair of point sets.

%%%%%%%%%%%%%%%%%%%%%%%%%%%%%%%%%%%%%%%%%%%%%%%%%%%%%%%%%%%%%%%%%%%%%%%%%%%%%%
\subsection{}

The problem of finding compatible triangulations can be made easier if we can
add extra points to the point sets, called \emph{Steiner} points. Steiner points placed inside (outside) the convex hull of a point set are named \emph{interior} (\emph{exterior}) Steiner points. Aichholzer et al.\ \cite{aichTCT} give a method of compatibly triangulating two
point sets with the introduction of a number of interior Steiner points equal to
the number of interior points minus two.  One might try methods of compatible
triangulation using exterior Steiner points, that is, additional points added
outside of the convex hull of the original point set. We show a method of doing
this that uses a number of Steiner points independent of the size of the point sets
to be triangulated.

\begin{obs}
\label{obvious}
Given a finite set $S$ of points in $\R^{2}$, coordinates for $\R^{2}$ may be chosen under which every point of $S$ has a distinct second coordinate:
$$ S = \{ \ (x_1, y_1), \cdots, (x_n, y_n) \suchthat y_{i} \neq y_{j} \ \text{
if } \ i \neq j \}. $$
\end{obs}

\begin{thm}
\label{t:twosteiner}
Given two point sets $S$ and $T$ with $|S| = |T| = n$, then $S$ and $T$ may be compatibly triangulated with the addition of two Steiner points to each set.
\end{thm}

\begin{proof}
We may assume both sets have the property described in Observation~\ref{obvious}. Order each set in terms of increasing second coordinate.  Then
$$ S = \{ \ p_1 = (x_1, y_1), \cdots, p_n = (x_n, y_n) \suchthat y_{i} > y_{j} \ \text{ if } \ i > j \}$$
and
$$T = \{ \ q_1 = (u_1, v_1), \cdots, q_n = (u_n, v_n)  \suchthat   v_{i} > v_{j} \ \text{ if } \ i > j\}.$$
Add a Steiner point $p_L$ far enough to the left of $S$ so that edges $p_i p_{i+1}$ (between consecutive points of $S$) and $p_L p_i$ (between $p_L$ and points of $S$) do not intersect pairwise. Similarly, add a Steiner point $p_R$ to the right of $S$ with the corresponding property; see Figure~\ref{f:twosteiner}.  Note that $p_L$ and $p_R$ exist  since placing them arbitrarily far away yields edges arbitrarily close to horizontal.  Add $q_L$ and $q_R$ to $T$ in the same manner.

\begin{figure}[h]
\includegraphics {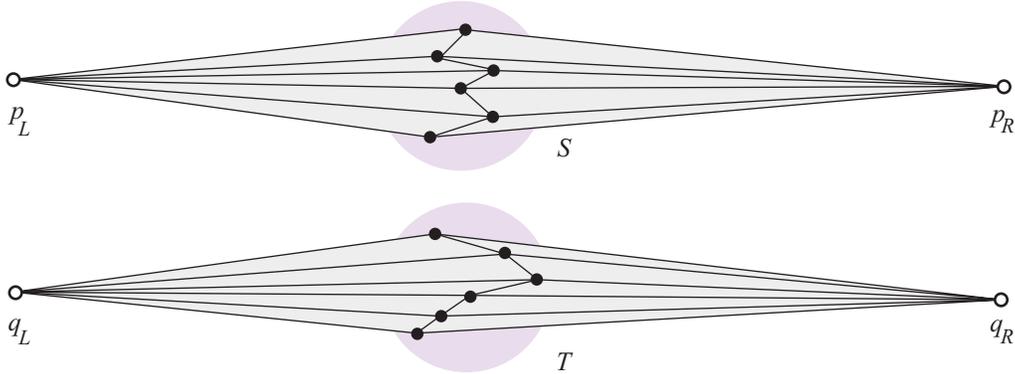}
\caption{Example of sufficiency of two exterior Steiner points.}
\label{f:twosteiner}
\end{figure}

Let $S_* = S \cup \{p_L, p_R\}$ and $T_* = T \cup \{q_L, q_R\}$.  By
construction, the edges $p_L p_i$ and $p_i p_R$ (connecting each Steiner point
to every point of the original set) together with the edges $p_i p_{i+1}$
(connecting vertically consecutive points of the original set) yield a
triangulation of $S_*$; a similar construction gives a triangulation to $T_*$.
The bijection $f(p_*) = q_*$ shows these triangulations to be compatible.
\end{proof}

The preceding theorem demonstrates the power of adding exterior Steiner points.
The downside is that the new Steiner points can be arbitrarily far away; a better solution would bound the maximum distance between points.
The \emph{radius} of a point set $S$, denoted by $r(S)$, is the radius of the smallest disk that contains all of $S$.
In Section~\ref{s:compatible}, a method of compatibly triangulating point sets will be presented where the Steiner points increase the radius of the point set by a fixed constant.

%%%%%%%%%%%%%%%%%%%%%%%%%%%%%%%%%%%%%%%%%%%%%%%%%%%%%%%%%%%%%%%%%%%%%%%%%%%%%%
%
%               Partitioning Points
%
%%%%%%%%%%%%%%%%%%%%%%%%%%%%%%%%%%%%%%%%%%%%%%%%%%%%%%%%%%%%%%%%%%%%%%%%%%%%%%
\section{Partitioning point sets with series-triangular graph embeddings}
\subsection{}

In order to construct compatible Steiner point triangulations, it is helpful to
break the problem into smaller pieces.  One way is to lay down the
Steiner points so that a triangulation of the Steiner points alone divides the
original point sets into smaller point sets in each triangle.  This only works
if the triangulations on the Steiner points are compatible and the same number
of points fall in corresponding triangles.  In this section we show this is
always possible.

Recall that Definition~\ref{d:comp2} states that compatibility is just a convex
hull preserving isomorphism.  Therefore, in order to add Steiner points that
can be compatibly triangulated, we view  these triangulations as two embeddings of the same graph in which the same vertices map to the convex hull in the same orientation.  The graphs used to perform this desired partition all have a common structure.

%In graph theory, a planar triangulation is a planar graph with a maximal edge
%set.  If a graph $G$ is a planar triangulation then every straight-line
%embedding of $G$ in the plane is a triangulation in the geometric sense with
%three convex hull points.  In this section, we will define a special class of
%planar triangulations and give conditions for two distinct straight-line
%embeddings to be compatible triangulations.  Later we will show that these
%triangulations can be used to partition point sets in nice ways.

% The following definition is intended to recall 'series-planar' graphs.
% I think this is a good name because like series-planar graphs, these
% graphs have a recursive definition that implies a procedure for building
% them from smaller graphs.  Also, the definition implies the existence
% of a series of graphs with this property, namely the intermediate steps in
% the construction.

\begin{defn}
A graph $G(V,E)$ is \emph{series-triangular} if
\begin{enumerate}
\item it is planar,
\item every embedding of $G$ in $\R^2$ is a triangulation, and
\item when $|V|>3$, there is a vertex $v\in V$ such that $G \setminus v$ is
series-triangular.
\end{enumerate}
\end{defn}

\begin{figure}[h]
\includegraphics {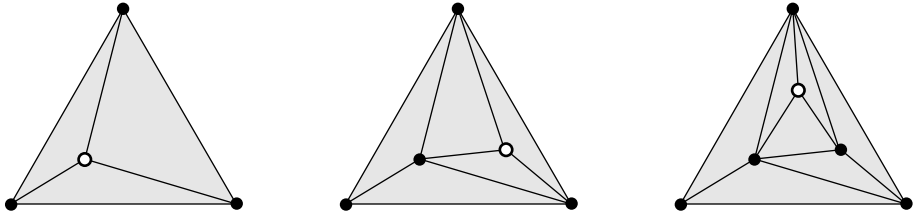}
\caption{Constructing series-triangular graphs one vertex at a time.}
\label{f:striangular}
\end{figure}

\noindent Figure~\ref{f:striangular} shows the construction of a series triangular graph, starting with a triangle, one vertex at a time.

The recursive definition above implies that there is a labeling $v_1, \ldots ,v_n$ of the vertices
of $G$ such that the subgraph $G_i$ induced on $\{v_1, \ldots ,v_i\}$ is series-triangular for $i \geq 3$; this is called an \emph{ordered labeling}.  In particular, if an ordered labeling is given, then it is possible to embed $G$ in $\R^2$ so that $\{v_1, v_2, v_3\}$ map to the convex hull
oriented clockwise.  An embedding that satisfies this property is called an
\emph{ordered embedding}.  An immediate consequence of Definition~\ref{d:comp2} is
the following useful lemma.

\begin{lem}
\label{l:stcomp}
Let $\phi,\psi$ be straight-line embeddings of a series-triangular graph $G$.
Then $\phi(G)$ and $\psi(G)$ are compatible triangulations under the bijection
$f(\phi(v))=\psi(v)$ if and only if $\phi$ and $\psi$ map the same vertices of
$G$ to the convex hull in the same orientation.
\end{lem}

In other words, for a given ordered labeling of the vertices of $G$, all ordered embeddings are compatible.  Thus, once an ordered labeling for $G$ (there are many) is chosen, the compatibility class of the ordered embedding is fixed.  In what follows, we always assume an ordered
labeling is given with $G$ and refer to the triangles in an ordered embedding of $G$ as simply the triangles of $G$.

%%%%%%%%%%%%%%%%%%%%%%%%%%%%%%%%%%%%%%%%%%%%%%%%%%%%%%%%%%%%%%%%%%%%%%%%%%%%%%
\subsection{}

Before analyzing issues with series-triangular graphs, we first examine the partition of point sets within a single triangle.
\begin{figure}[h]
\includegraphics {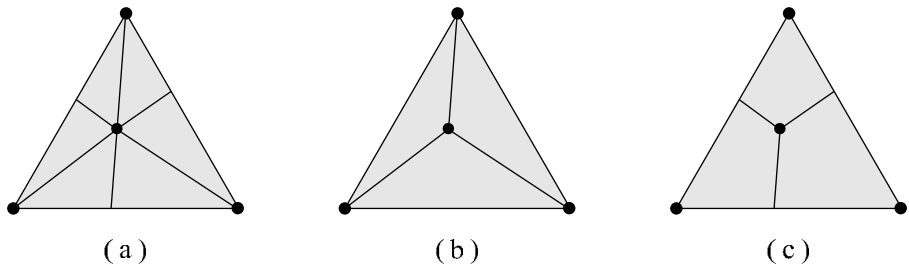}
\caption{(a) Subdivision, (b) trisection, and (c) Y-section of a triangle.}
\label{f:tsub}
\end{figure}
\begin{defn}
Let $p$ be an interior point of a triangle $T \subset \R^2$.  The three lines
passing through $p$ and each vertex of $T$ \emph{subdivides} $T$ into six
triangles.  The \emph{trisection} of $T$ at $p$ is obtained from the three line segments from each vertex of $T$ to $p$.  The three halflines emanating from $p$ directed away from the three vertices of $T$ give the \emph{Y-section} of $T$ at $p$, denoted as $Y(p)$; see Figure~\ref{f:tsub}.
\end{defn}

If a Steiner point $d$ is placed inside $T$, the trisection of $T$ at $d$ divides $T$ into three adjacent triangles.  Ideally, we would like to have the freedom to place $d$ in such a way that the number of points in each of the triangles of the trisection is prescribed.  The following shows  this is always possible.

\begin{lem}
\label{partitioncount}
Let $S=\{p_1,\ldots ,p_n\}$ be a set of points inside a triangle
$T=\tri(a,b,c)$ satisfying the following conditions:
\begin{enumerate}
\item The points $S\cup \{a,b,c\}$ are in general position.
\item If $A$ is the arrangement of all lines passing through
pairs of points $(t,s)$ where $t\in \{a,b,c\}$ and $s\in S$ then the lines of
$A$ have no three-way intersection inside $T$.
\end{enumerate}
Then the set $Y(p_1) \cup \cdots \cup Y(p_n)$ divides $T$ into $\binom{n+2}{2}$
regions.
\end{lem}

\begin{figure}[h]
\includegraphics {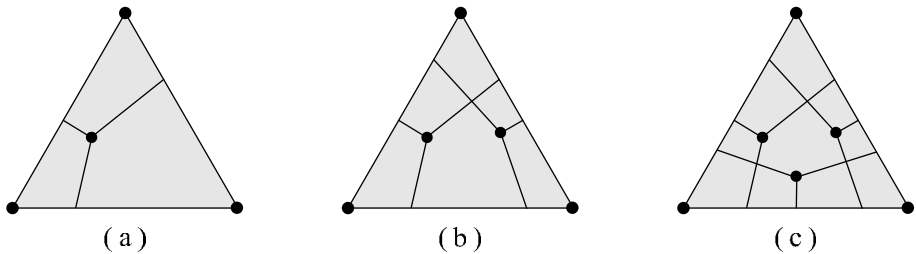}
\caption{The Y-sections of (a) one, (b) two and (c) three points.}
\label{f:y123}
\end{figure}

\begin{proof}
We construct this inductively, starting with $T$ and adding a point of $S$ at each step.  It is clear from the definition that every pair of distinct $Y$-sections
intersect at  exactly one point.  It follows from the assumption that no
three $Y$-sections intersect at a point.  Thus, $k$ distinct $Y$-sections result in $k-1$ intersections.  Each additional $Y$-section divides the region of its center point into three parts and divides each of the $k-1$ other regions into two parts, one for each line crossed; Figure~\ref{f:y123} shows some examples. Thus
$$\sum_{i=0}^n (i+1) =  \binom{n+2}{2}$$
is the total number of regions for $n$ points.
\end{proof}

\begin{lem}
\label{wheretoputit}
Let $S=\{p_1,\ldots ,p_n\}$ be a set of points inside a triangle $T$ satisfying the conditions of Lemma~\ref{partitioncount}.  Let $x,y,z$ be positive integers such that $x+y+z=n$.  A Steiner point $d$ can be placed in $T$ such that the three triangles formed by the trisection of $T$ at $d$ contain $x,y,z$ interior points respectively.
\end{lem}

\begin{proof}
There are $\binom{n+2}{2}$ ways to express the number $n$ as the ordered sum of
three nonnegative integers $x, y, z$.  We show that each of the $\binom{n+2}{2}$
regions defined by the $Y$-sections of $S$  achieves a unique ordered sum
$x+y+z=n$ when a Steiner point is added.  Suppose for contradiction that there
are Steiner points $p_1$ and $p_2$ in two distinct regions that achieve the same partition $x + y + z = n$ of points. Then  $p_1$ (similarly $p_2$) trisects $T$ into triangles
$A_x,A_y,$ and $A_z$ (similarly $B_x,B_y$, and $B_z$) with $x,y,$ and $z$ interior points respectively; see Figures~\ref{f:intregion}(a) and (b).

\begin{figure}[h]
\includegraphics {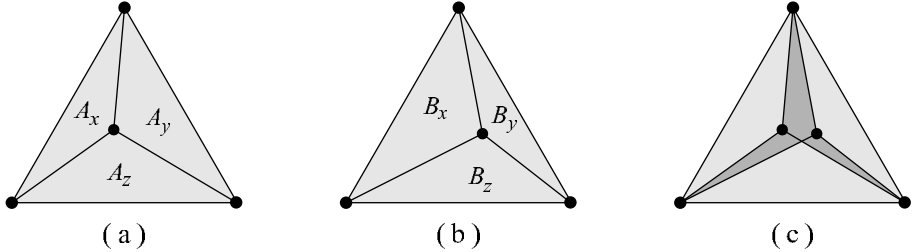}
\caption{Trisections and intersections of regions.}
\label{f:intregion}
\end{figure}

Let $M = (A_x \oplus B_x) \, \cup \, (A_y \oplus B_y) \, \cup \, (A_z \oplus B_z)$, the shaded region of Figure~\ref{f:intregion}(c) ---  the symbol $\oplus$ denotes the symmetric difference between the two sets.  Because $p_1$ and $p_2$ are in different regions, the line segment connecting $p_1$ to $p_2$ intersects an edge of a $Y$-section, and so there must be some point of
$S$ in $M$.  Without loss of generality, assume $A_x \subset B_x$ and $B_y \subset A_y$.  This implies $S \cap A_z = S \cap B_z$ and thus $S \cap M$ is empty, a contradiction.
\end{proof}

\begin{rem}
It follows from the proof that there is a nonempty, open region in $T$, any point of which will achieve the desired result.
\end{rem}

%%%%%%%%%%%%%%%%%%%%%%%%%%%%%%%%%%%%%%%%%%%%%%%%%%%%%%%%%%%%%%%%%%%%%%%%%%%%%%
\subsection{}

We now show how a series-triangular graph can be embedded in the plane so that the triangular faces of the embedding partition a set of points nicely.

\begin{thm}
\label{t:partition} Let $S$ be an $n$ point set in $\R^2$ and
$G(V,E)$ be a series-triangular graph, and let $k$ be odd.  If
$\{t_1,\ldots ,t_k\}$ are the triangles of $G$ and $n_1 + \cdots +
n_k$ is a partition of $n$, then there is an ordered embedding
$\phi(G)$ with the property that exactly $n_i$ points of $S$ lie
in the triangle $t_i$.
\end{thm}

\begin{proof}
We construct $\phi$ by induction on $|V| = \frac{k+5}{2}$. The
base case when $|V|=3$ and $k=1$ is trivial because any triangle
enclosing all of $S$ suffices.  We choose this triangle so that it
satisfies the conditions of Lemma~\ref{partitioncount}.  Now,
assume $k \geq 3$ and let $n' = n_{k-2}+n_{k-1}+n_{k}$. As $G$ is
series-triangular, let $v$ be the vertex such that when removed
along with all edges touching it, the resulting graph, called
$G'$, is series-triangular. $G'$ has one less vertex and two less
triangles, so by induction we can find an embedding $\phi'(G')$
such that $n_1,n_2,\ldots, n_{k-3}, n'$ points of $S$ lie in the
respective triangles $t_1',t_2',\ldots,t_{k-3}',t_{k-2}'$ of
$\phi'(G')$. Further assume by induction that so far the
conditions of Lemma~\ref{partitioncount} are met. Now,
Lemma~\ref{wheretoputit} ensures that there is a non-empty open
region inside $t_{k-2}'$, any point of which, when connected to
the vertices of $t_{k-2}'$, will divide the triangle into three
smaller triangles having exactly $n_{k-2}$, $n_{k-1}$ and $n_{k}$
of the $n'$ points of $t_{k-2}'\cap S$. From this region we may
pick a point $p$ so that when we define $\phi$ by extending
$\phi'$ to include $\phi(v) = p$, the conditions of
Lemma~\ref{partitioncount} are satisfied. The three new triangles
formed by the addition of $\phi(v)$ are the triangles $t_{k-2},
t_{k-1}, t_{k}$ of the embedding $\phi(G)$.
\end{proof}

% Next, assume that the
%theorem holds for $|V|=k-1$ and that each triangle in
%$\phi(G_{k-1})$ satisfies the conditions of
%Lemma~\ref{partitioncount}, where $G_{k-1}$ is .
%Choose $a, b, c$ so that $t_a, t_b, t_c$ are the three triangles created
%when $v_{i+1}$ is embedded.  Replace $t_a, t_b, t_c$ with a new
%triangle $t'$ and replace $n_a, n_b, n_c$ with $n_a + n_b + n_c$.  By
%induction $\phi(G_i)$ exists, and Lemma~\ref{wheretoputit} ensures that there is a
%region where $v_{i+1}$ can be embedded in $t'$ so that triangles $t_a, t_b, t_c$ contain $n_a, n_b, n_c$ points of $S$, respectively.  Because this region is nonempty and open, there is room to perturb $\phi(v_{i+1})$ in order to satisfy the conditions of Lemma~\ref{partitioncount}.

\noindent In a sense, this construction stretches the triangulation over the point set controlling how many points land in each triangle.  It is not hard to see that this partitioning theorem can be modified so that the number of points in the unbounded face can also be specified.  This is a  trivial exercise that we omit here.

%%%%%%%%%%%%%%%%%%%%%%%%%%%%%%%%%%%%%%%%%%%%%%%%%%%%%%%%%%%%%%%%%%%%%%%%%%%%%%
%
%                Compatibility
%
%%%%%%%%%%%%%%%%%%%%%%%%%%%%%%%%%%%%%%%%%%%%%%%%%%%%%%%%%%%%%%%%%%%%%%%%%%%%%%
\section{Compatible Steiner point triangulations}
\label{s:compatible}
\subsection{}

We will now show that the partitioning theorem in the previous section can be
used to triangulate point sets with Steiner points.  This method will embed a series-triangular graph around each point set partitioning them so each triangle receives one point.  The new points introduced will be the Steiner points and the compatible triangulation of those points will be extended to a compatible triangulation of the entire point sets.

A partition of
points by  a series-triangular graph embedding encloses the entire point set in
a triangle.  The smallest triangle containing the disk of radius $r(S)$ has
radius $2r(S)$, so the radius of the point set only increases by a factor of $2$ if we add the points needed to partition it.

%we may want a figure here.  it would be a circle inscribed in an equilateral
% triangle inscribed in another circle.  The triangle would be subdivided into
% 6 30-60-90 triangles demonstrating the radius doubling because the short edge
% is the inner radius and the hypotenuse is the outer radius.

\begin{thm}
\label{t:kovertwo}
Given point sets $S,T$ with $|S|=|T|=n$, it is possible to compatibly
triangulate $S$ and $T$ with the addition of at most $\frac{n}{2}+3$ Steiner
points.
\end{thm}

\begin{proof}
Let $G$ be any series-triangular graph with at least $n$ bounded triangles.
Use Theorem~\ref{t:partition} to obtain an ordered embedding of $G$ in
$\R^2$ so that no two points of $S$ share a triangle; repeat this
process for $T$.  By Lemma~\ref{l:stcomp}, the embeddings are compatible triangulations.
Choose the partitions so that compatible triangles contain the same number of vertices ($0$
or $1$ in this case).  Because each pair of compatible triangles yields a
compatible triangulation of its interior points, these triangulations are easily
extended to compatible triangulations of the entire point sets.  Since any planar triangulation on $k$ vertices has $2k-5$ bounded triangles, we
can choose $G$ such that $|V|=\lceil\frac{n+5}{2}\rceil< \frac{n}{2}+3$.
\end{proof}

%%%%%%%%%%%%%%%%%%%%%%%%%%%%%%%%%%%%%%%%%%%%%%%%%%%%%%%%%%%%%%%%%%%%%%%%%%%%%%
\subsection{}

Aichholzer et al.\ \cite{aichTCT} show that for point sets with only three
internal points, there exists a compatible triangulation as long as their
convex hulls are the same size.  Moreover, this result holds even if the
bijection between their convex hulls is prescribed by cyclic rotation.  An
immediate consequence of this fact is that the proof of the
preceding theorem can be modified to place three points in each triangle rather than one.  Thus
we get the following:

\begin{cor}
Given point sets $S,T$ with $|S|=|T|=n$, it is possible to compatibly
triangulate $S$ and $T$ with the addition of at most $\frac{n}{6}+3$ Steiner
points.
\end{cor}

Computer simulations in \cite{aichTCT} have shown that all sets of $8$ points of the same convex hull size yield a compatible triangulation even if the
bijection between extreme points is fixed by cyclic rotation.  Thus, this result can again be refined by placing  $5$ points in each triangle.

\begin{cor}
Given point sets $S,T$ with $|S|=|T|=n$, it is possible to compatibly
triangulate $S$ and $T$ with the addition of at most $\frac{n}{10}+3$ Steiner
points.
\end{cor}

These corollaries demonstrate the expandability of the series-triangular
partition method.  As larger point sets are shown to yield compatible
triangulations with extreme points prescribed by cyclic rotation, the number of
Steiner points needed to compatibly triangulate goes down.  Our methods give a
framework for directly extending results on small point sets to arbitrary point
sets.

%%%%%%%%%%%%%%%%%%%%%%%%%%%%%%%%%%%%%%%%%%%%%%%%%%%%%%%%%%%%%%%%%%%%%%%%%%%%%%
%
%                d-way Compatibility
%
%%%%%%%%%%%%%%%%%%%%%%%%%%%%%%%%%%%%%%%%%%%%%%%%%%%%%%%%%%%%%%%%%%%%%%%%%%%%%%
\section{$d$-way Compatible Triangulations}
\subsection{}

We pose a more general question in which there are not just two point sets but
rather $d$ point sets $S_1,\ldots,S_d$.  Thus, instead of a single bijection, a set of bijections $f_1,\ldots,f_{d-1}$ are needed with the maps $f_i: S_i \to S_{i+1}$ mapping
triangulations to triangulations compatibly.  An important application of
compatible triangulations comes from the problem of \emph{morphing} computer
graphics.  Surazzhsky and Gotsman \cite{morph} show that if a pair of compatible
triangulations are given, then it is possible to linearly morph one into the
other without any triangle edges intersecting.  Suppose one desires to morph a
triangulated point set $S_1$ to a point set $S_2$ and then again to third point set
$S_3$ and so on to $S_d$.  In such a case, a $d$-way compatible triangulation is
necessary.  Indeed, previous methods for finding compatible Steiner triangulations would
require adding new Steiner points for each morph thus making the number of
Steiner points dependent on $d$.  However, it is shown below that our method for
compatible Steiner triangulation of two point sets extends to $d$-way compatible
triangulations so that the number of Steiner points stays linear and is
independent of $d$.

\begin{defn}
\label{d:dway}
Given point sets $S_1,\ldots,S_d$, with $|S_i|=|S_j|$ for all $i,j$, and
triangulations $\T_1,\ldots,\T_d$, then $\{\T_1,\ldots,\T_d\}$ is \emph{$d$-way
compatible} if $\T_i$ and $\T_j$ are compatible for all $i,j$.
\end{defn}

\begin{figure}[h]
\includegraphics {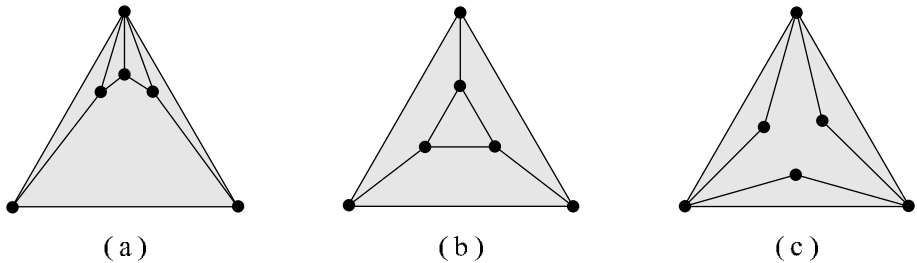}
\caption{Three point sets and their forced edges without compatible triangulations.}
\label{f:threewaybreakdown}
\end{figure}

As stated earlier, it is not known whether Steiner points are necessary for a compatible triangulation of two point
sets.  The open conjecture is that the compatible triangulation always exists
when the point sets are the same size and have the same number of extreme
points.  The corresponding conjecture for $d$-way compatible triangulation is
false even for the case when $d=3$.  Figure~\ref{f:threewaybreakdown} depicts a
simple counterexample.

\noindent In the figure above, the point sets are drawn along with the forced edges --- edges which appear in every triangulation of that point set.  It is easy to see that set (a) forces one extreme point to be connected to every interior point and (b) forces the interior points to form a triangle.  If both of these demands are met on point set (c), then two edges must cross.

%%%%%%%%%%%%%%%%%%%%%%%%%%%%%%%%%%%%%%%%%%%%%%%%%%%%%%%%%%%%%%%%%%%%%%%%%%%%%%
\subsection{}

The example above shows that even in very simple cases, Steiner points are
necessary for a $d$-way compatible triangulation.  Surprisingly, the number of
Steiner points needed for each point set is independent of $d$.

\begin{thm}
Given point sets $S_1,\ldots,S_d$ with $|S_i|=n$, it is possible to
$d$-way compatibly triangulate $S_1,\ldots,S_d$ with the addition of at most
$\frac{n}{2} + 3$ Steiner points to each set.
\end{thm}

\begin{proof}
Construct any series-triangular graph $G(V,E)$ with at least $n$ triangles.  By
Theorem~\ref{t:partition}, it is possible to embed $G$ in $\R^2$ for each $S_i$ so that one
point of $S_i$ falls in each triangle.  There may be one empty triangle if $n$
is even but this is inconsequential as long as we choose the same triangle to
remain empty in every embedding.  All $d$ embeddings $\phi_1,\ldots,\phi_d$ are compatible by
Lemma~\ref{l:stcomp}.

Let $\{t_1,\ldots,t_n\}$ be the triangles of $G$.  For each
$p\in S_i$ and $q\in S_{i+1}$, let $f_i(p)=q$ if and only if $p$ is inside
triangle $\phi_i(t_j)$ and $q$ in $\phi_{i+1}(t_j)$ for some $j$.  Extend
the bijections to include the Steiner points by letting
$f_i(\phi_i(v))=\phi_{i+1}(v)$.  All remaining edges are forced and compatible;
they are simply the edges connecting each point in the original sets to the
three vertices of the triangle that encloses it.  As in the proof of
Theorem~\ref{t:kovertwo}, one can construct $G$ so that
$|V|\le\frac{n}{2}+3$.  The Steiner points for each point set $S_i$ are
$\phi_i(V)$; thus the number of Steiner points required for each set is also
at most $\frac{n}{2}+3$.
\end{proof}

\begin{figure}[h]
\includegraphics {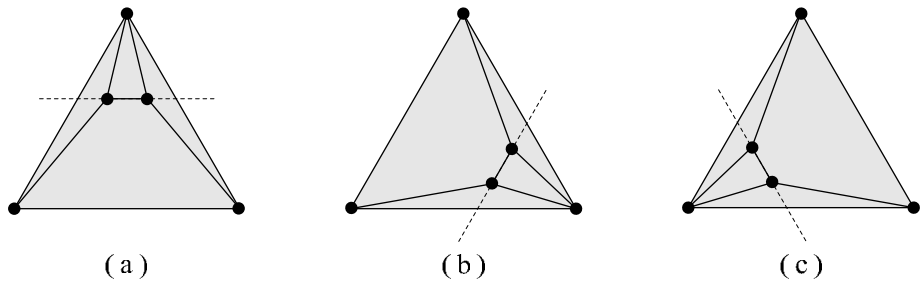}
\caption{Three point sets in Steiner point triangles and their forced edges.}
\label{f:twopointbreakdown}
\end{figure}

Unlike the case where $d=2$, it is not possible to extend this method by placing
more points inside each triangle of the embedded graph.  We note that if even
two points go into any one triangle, it may be impossible to the extend the
compatible triangulation of the Steiner points to a compatible triangulation of
the entire points sets.  Figure~\ref{f:twopointbreakdown} gives a simple example
where this is the case.  Note that at least one of the extreme points must have degree $3$; this is not possible however since each of the extreme points has four forced
edges in at least one embedding.

\begin{ack}
We thank John Mugno and Rachel Ward for helpful discussions.
\end{ack}

%%%%%%%%%%%%%%%%%%%%%%%%%%%%%%%%%%%%%%%%%%%%%%%%%%%%%%%%%%%%%%%%%%%%%%%%%%%%%%
%
%                  REFERENCES
%
%%%%%%%%%%%%%%%%%%%%%%%%%%%%%%%%%%%%%%%%%%%%%%%%%%%%%%%%%%%%%%%%%%%%%%%%%%%%%%
\bibliographystyle{amsplain}

\end{document}